\documentclass[aps,prl,twocolumn,superscriptaddress,showpacs]{revtex4-1}
\usepackage{graphicx,psfrag}
\graphicspath{{/home/user/fig/publi/muon/}}
\DeclareGraphicsExtensions{.ps, .eps}

\begin{document}

\title{Evidence for an exotic magnetic transition in the triangular spin system 
FeGa$_2$S$_4$} 

\author{P.~\surname{Dalmas de R\'eotier}}
\author {A.~Yaouanc}
\affiliation{Institut Nanosciences et Cryog\'enie, SPSMS, CEA and University Joseph Fourier, F-38054 Grenoble, France}
\author{D.~E.~MacLaughlin}
\author{Songrui~Zhao}
\altaffiliation{Now at Department of Electrical and Computer Engineering, McGill University, Montreal, Quebec, Canada H3A 2A7}
\affiliation{Department of Physics and Astronomy, University of California, Riverside, California 92521, USA}
\author{T.~Higo}
\author{S.~Nakatsuji}
\author{Y.~Nambu}
\affiliation{Institute for Solid State Physics, University of Tokyo, Kashiwa 277-8581, Japan}
\author {C.~Marin}
\author {G.~Lapertot}
\affiliation{Institut Nanosciences et Cryog\'enie, SPSMS, CEA and University Joseph Fourier, F-38054 Grenoble, France}
\author{A.~Amato}
\author{C.~Baines}
\affiliation{Laboratory for Muon-Spin Spectroscopy, 
Paul Scherrer Institute, 5232 Villigen-PSI, Switzerland}

\date{\today}

\begin{abstract}

We report positive muon spin relaxation measurements on the triangular lattice magnetic system FeGa$_2$S$_4$. A magnetic transition not previously detected by specific heat and magnetic susceptibility measurements is found in zero field at $T^* \simeq 30$~K\@.
It is observed through the temperature dependencies of the signal amplitude and the spin-lattice relaxation rate. This transition is therefore not a conventional magnetic phase transition. 
Since persistent spin dynamics is observed down to 0.1~K, the ground state cannot be of the canonical spin-glass type, which could be suggested from hysteresis effects in the bulk susceptibility below $T_f \simeq 16$~K\@. These results are compared to those found for the isomorph NiGa$_2$S$_4$. It is argued that the fate of the transition, which has been interpreted in terms of the $Z_2$ topological transition in this latter system, is probably different in FeGa$_2$S$_4$.

\end{abstract}

\pacs{75.40.-s, 75.40.Gb, 76.75.+i}
\maketitle

The search for new states of matter is at the forefront of condensed matter research%
, and the geometrically frustrated magnetic systems provide a fruitful playground \cite{Ramirez01,Lee08,Balents10}. One of the main goals is to find and characterize magnetic systems for which long-range magnetic order is absent at low temperature despite strong exchange interactions. Two-dimensional Heisenberg antiferromagnets are good candidates since magnetic order can only occur at temperature $T = 0$. In the case of the 
equilateral triangular lattice, it is now believed that the system would exhibit $120^\circ$ spin order \cite{Huse88,Bernu92,Capriotti99}.
However, symmetry breaking energy terms and exchange interactions between the layers or inside the layers beyond the nearest-neighbor may lead to magnetic order at finite temperature or quantum spin-disordered ground states \cite{Diep04}. Only a few two-dimensional triangular-lattice systems are believed not to display magnetic long-range order: two organic and two inorganic compounds \cite{Nakatsuji10}. While recent interesting results have been reported \cite{Pratt11} the restricted amount of organic material that can be produced imposes serious limitations on the measurements that can be performed, precluding e.g. inelastic neutron scattering experiments. This is not the case for the inorganic transition metal sulfides NiGa$_2$S$_4$ and FeGa$_2$S$_4$, which are available in large quantities. These two insulators, which crystallize in the space group $P{\bar 3}m1$, consist of magnetic Ni$^{2+}$ and Fe$^{2+}$ layers, respectively, characterized by strong intralayer magnetic bonds in comparison to very weak
interlayer couplings. The Ni and Fe ion sites form an equilateral triangular lattice in each layer. 

While NiGa$_2$S$_4$ has been studied with a wide range of macroscopic and microscopic experimental techniques \cite{Nakatsuji05,Nakatsuji07a,Takeya08,Yamaguchi08,Yaouanc08,MacLaughlin08,Nambu09,Dalmas09}, fewer studies including X-ray diffraction, magnetic susceptibility, resistivity, and specific heat measurements have 
been reported for FeGa$_2$S$_4$ \cite{Nakatsuji07b,Tomita09}. The available data suggest that the two magnetic systems are similar. They have comparable lattice parameters. Their Curie-Weiss temperatures $\theta_{CW}$ are quite large, 80 and 160~K for NiGa$_2$S$_4$ and FeGa$_2$S$_4$
respectively. Upon cooling below 100~K, the dc susceptibility $\chi_\mathrm{dc}$ of the two compounds develops weak easy-plane anisotropy (which is larger for FeGa$_2$S$_4$) 
but no in-plane anisotropy. $\chi_\mathrm{dc}(T)$ shows a bifurcation between zero-field cooled (ZFC) and field-cooled (FC) data at 8.5--9~K and 16~K (i.e. $\simeq \theta_{CW}/10$) for NiGa$_2$S$_4$ and FeGa$_2$S$_4$, respectively. The magnetic contributions to the specific heat $C_{M}(T)$ of the two compounds are field insensitive with a $T^2$ dependence at low temperature, and they match one another after appropriate rescaling \cite{Nambu08}. While $C_{M}/T$ is negligible for $T \rightarrow 0$~K in NiGa$_2$S$_4$, an appreciable value is measured for FeGa$_2$S$_4$. Finally $C_{M}/T$ exhibits an unusual double-peak structure: a first rounded maximum at $\approx$~10~K for the two systems, and a second broad maximum centered around 100~K and 60~K for NiGa$_2$S$_4$ and FeGa$_2$S$_4$, respectively. 

The microscopic magnetic properties of NiGa$_2$S$_4$ are experimentally well established. Neutron scattering experiments \cite{Nakatsuji05} show that 
NiGa$_2$S$_4$ displays incommensurate quasi-static short-range magnetic correlations. These correlations manifest themselves in a spontaneous field below $T^*\simeq 9$~K in muon spin rotation and relaxation ($\mu$SR) experiments \cite{Yaouanc08,MacLaughlin08}. 
Concerning 
temperatures above  $T^*$: while the longitudinal-field $\mu$SR relaxation function is exponential at high temperature, it becomes 
sub-exponential or ``stretched exponential'' for
$T \to T^*$ and below. This shape
indicates the presence of
spatial inhomogeneity in the relaxation rate. On the premise 
that topological $Z_2$ vortices manifest themselves in the vicinity of the temperature where $C_{M}(T)$ displays a broad maximum \cite{Kawamura84}, the transition at $T^*$ was tentatively associated with the dissociation of these vortices \cite{Yaouanc08}. In addition, the stretched exponential relaxation just above $T^*$ was interpreted as the signature of vortex unbinding, which is expected to give rise to magnetic disorder.

We are aware of only one investigation of FeGa$_2$S$_4$ by a microscopic technique. From $^{57}$Fe M\"ossbauer spectroscopy, a spontaneous hyperfine field has been observed at low temperature with an order-parameter-like temperature dependence \cite{Myoung08,Myoung10}. An antiferromagnetic transition has been inferred from these data with a N\'eel temperature of 33~K. In this report we present $\mu$SR measurements which are aimed at further characterising the nature of the transition. We also find a magnetic transition
with $T^* \simeq$ 30\,K $\simeq \theta_{CW}/5$ rather than 9\,K $\simeq \theta_{CW}/10$ in the nickel counterpart.
Persistent spin dynamics are observed in both compounds down to the lowest temperature of the measurements, i.e., 0.1~K or less \cite{[{For NiGa$_2$S$_4$, see }]MacLaughlin10}. 
Nevertheless, the two systems differ sharply in 
some respects. In contrast to NiGa$_2$S$_4$ no spontaneous field is detected in
FeGa$_2$S$_4$, owing to a broader field distribution.
In the latter compound no anomaly is observed in $C_{M}(T)/T$ or $\chi_{\mathrm{dc}}(T)$ at $T^*$, whereas in NiGa$_2$S$_4$ a weak cusp is seen in $\chi_{\mathrm{dc}}(T)$ at low fields. 
In addition none of the temperatures of the two $C_M$ maxima in the two systems scales with $T^*$.
The observation of a magnetic transition in FeGa$_2$S$_4$ while no signature for it is detected in specific heat and susceptibility data points to its exotic nature.

Data were obtained from $\mu$SR measurements carried out at the General Purpose Surface-Muon (GPS) ($2.8~\mathrm{K} \le T \le 160$~K) and the Low Temperature Facility (LTF) ($0.10~\mathrm{K} \le T \le 3.5$~K) instruments of the Swiss Muon Source (S$\mu$S), Paul Scherrer Institute (PSI), Villigen, Switzerland. Most of the spectra were recorded in zero field, with additional spectra taken with an applied external longitudinal field. Measurements were mostly performed on a powder sample prepared in Tokyo. A few spectra were also recorded on a powder prepared in Grenoble, giving results consistent with a transition at $T^*$. 

$\mu$SR techniques and their application to the study of magnetic materials are described elsewhere \cite{Dalmas97,Dalmas04,Yaouanc11}.
A time-differential $\mu$SR asymmetry spectrum is given by $a_{0}\eta P^{\rm exp}_Z(t)$, 
where $a_{0}$ is the spectrometer-dependent initial asymmetry, $P^{\rm exp}_Z(t)$ is the muon spin polarization function 
[$P^{\rm exp}_Z(t{=}0) = 1$] \cite{Yaouanc11}, and $\eta$ ($0 \le \eta \le 1$) characterizes the so-called missing asymmetry
due to rapid relaxation within the spectrometer dead time $t_{\rm dt}$
($\approx$~5~ns at PSI)\@. This is typically
caused by a broad distribution of static fields arising from frozen magnetism. 
The remaining asymmetry arises from the muon spin component parallel to the static field, and its relaxation is due solely to dynamic processes (spin-lattice relaxation) \cite{Yaouanc11}. In the following $a_{0}\eta$ will be called the effective initial asymmetry and $a_{0}\eta P^{\rm exp}_Z(t)$ the asymmetry.

\begin{figure}
\includegraphics[scale=0.75]{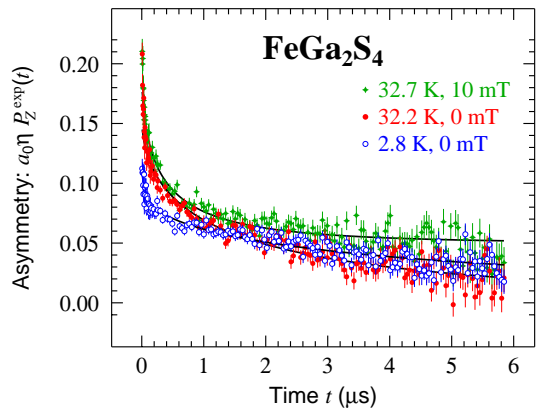}
\caption{(Color online) Examples of zero- and longitudinal-field $\mu$SR spectra recorded on a powder sample of FeGa$_2$S$_4$. Solid curves: results of fits as explained in the text. The reduced signal amplitude, i.e., the reduced effective initial asymmetry, for the $2.8$~K spectrum is clearly seen graphically. Note that these data result from the weighted subtraction of counts recorded in two opposite detectors, implying that their shape and amplitude are not dependent of any fit parameter.}
\label{muon_spectra}
\end{figure}

Exploratory measurements at the GPS spectrometer
indicated that the field history may have to be taken into account for $T < 60$~K\@. Hence, we first report zero-field measurements, for which no field was present when cooling down the sample below $60$~K\@. In contrast to NiGa$_2$S$_4$, no early-time oscillation could be reliably
resolved in the zero-field spectra at any temperature down to $0.1$~K\@. Thus we have not detected any spontaneous mean field at the muon site. This and the reduced low-temperature value of $\eta$, discussed below, indicate a broad distribution of fields. A lower bound for the width of the distribution at low temperature is of order $\Delta B  \approx 1/(\gamma_\mu t_{\rm dt}) \simeq$ 0.2~T ($\gamma_\mu = 851.6$\,Mrad s$^{-1}$T$^{-1}$ is the  muon gyromagnetic ratio). All the spectra can reasonably be described, except at short times ($t \lesssim $ 50~ns) near $T^*$ where the static fields become smaller, by the stretched exponential relaxation function
\begin{equation}
P^{\rm exp}_Z(t) = \exp\left [- {\left (\lambda_Z t \right )}^\beta \right ],
\label{fit_1}
\end{equation}
where $\lambda_Z$ is the spin-lattice relaxation rate and $\beta$ with $ 0< \beta\leq 1$ is the stretching power. Two typical zero-field spectra are shown in Fig.~\ref{muon_spectra}. 
We recall that a stretched exponential relaxation function accounts for a continuous distribution of relaxation rates. But as shown for La$_{1-x}$Ca$_x$MnO$_3$ 
this may not be justified \cite{Heffner00}, and a multi-site model might be more appropriate. Such a description would require so many parameters (two parameters per site, i.e., a weight and an exponential relaxation rate) that the physics would be obscured
, and we shall not attempt it. 

Focusing first on the spectra recorded at the GPS, and assuming $\beta$ to be a free parameter, we find $0.4 \lesssim \beta \lesssim 0.6$ for $2.8 \le T \le 25$~K and $0.3 < \beta < 0.4$ for $25 < T \le 35$~K\@. The value
above $40$~K increases smoothly to reach $\beta =1$ at $\approx$~60~K\@.
The final fits
were carried out setting $\beta$ = 1/2 somewhat arbitrarily for $T \le 40$~K and leaving it free above that temperature. 
Indeed, when $\beta$ approaches 0.3, 
$a_0\eta$ becomes unrealistically large. This points out the inadequacy of the fitting function for $t \lesssim 50$~ns. 
In contrast, setting $\beta$ = 1/2 
yields a value for $a_0\eta$ that fairly reflects the observed signal at short times. As shown in Fig.~\ref{muon_spectra} the resulting description of the spectra is reasonable.

It is more difficult to analyze the spectra recorded at the LTF than at the GPS because
a fraction of the muon beam stops in the silver cold finger of the dilution refrigerator. This fraction is traditionally determined from a calibration measurement performed above $T^*$ \cite{Yaouanc11}. Here this measurement cannot be done since $T^*$ is higher than the maximum temperature available at the LTF\@. To circumvent this difficulty 
spectra with overlapping temperatures 
were recorded at both the GPS and LTF; the $\lambda_Z$ values extracted from the GPS spectra
were used to analyze the LTF data, with the effective initial asymmetry being the only free physical parameter in the fitting procedure. With this method a reliable $a_{0}\eta$ value was obtained 
from the LTF spectra.
For temperatures at which only LTF spectra are available, this value 
was fixed during the fits. In addition, following the GPS results, $\beta$ = 1/2 
was assumed down to $0.1$~K, leaving $\lambda_Z$ as the only free physical parameter. 

\begin{figure}
\includegraphics[scale=0.75]{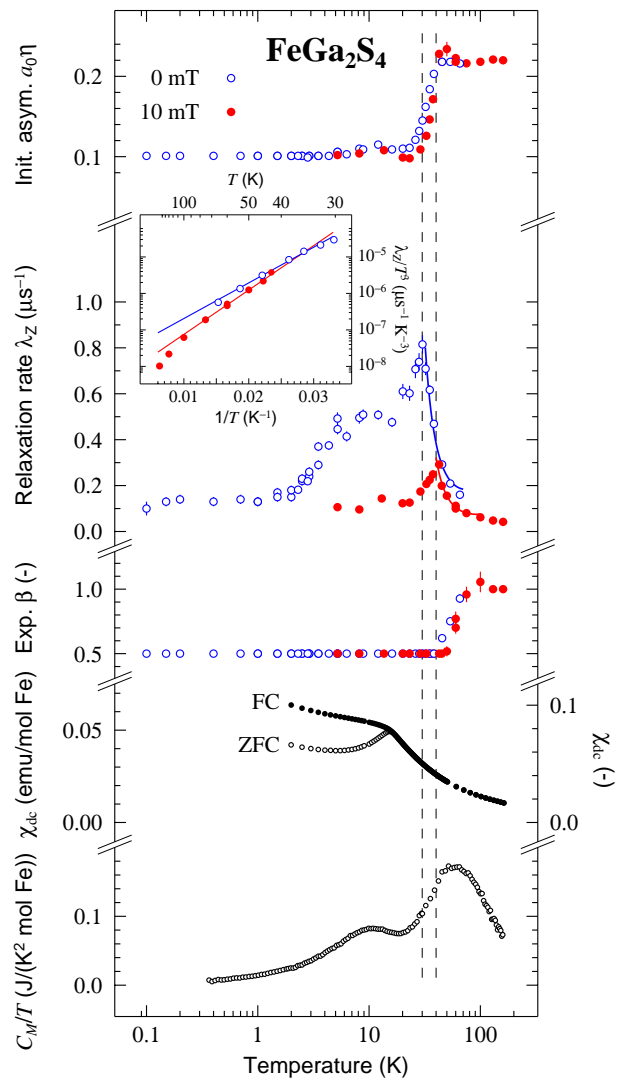}
\caption{(color online) The three upper panels show the temperature dependencies of the effective initial asymmetry $a_{0}\eta$, the spin-lattice relaxation rate $\lambda_Z$, and the stretched-exponential power $\beta$ measured in a powder sample of FeGa$_2$S$_4$ in zero 
or 10~mT longitudinal fields. 
The solid curves for $\lambda_Z(T)$ above $T^*$ result from a fit to a model explained in the main text. The adequacy of the model can be judged in the 
insert which displays $\lambda_Z/T^3$ versus $1/T$.
The temperature dependencies of the susceptibility $\chi_\mathrm{dc}$ given in CGS or SI units and the ratio $C_M/T$ of the magnetic specific heat over the temperature \cite{Nakatsuji07b} are shown in the lower two panels. The susceptibility has been measured in an external field of 0.1~T after a ZFC or FC protocol. For reference dashed lines are plotted at 30 and 40~K.}
\label{muon_results}
\end{figure}

Figure~\ref{muon_results} gives the temperature dependencies of $a_{0}\eta$, $\lambda_Z$ and $\beta$. 
We note the following three points. (i) The strong anomalies in $a_{0}\eta(T)$ and $\lambda_Z(T)$ at $T^* \simeq 30$~K point to a magnetic transition at $T^*$, independent of the details of the fitting procedure. This is a new evidence, after $^{57}$Fe M\"ossbauer spectroscopy \cite{Myoung08}, of such a transition in FeGa$_2$S$_4$. Its width is relatively large since the decrease of $a_{0}\eta$ with cooling begins at $40$~K and ends only at $\approx 
25$~K\@. The increase of $\lambda_Z$ by a factor 5 on cooling from 65~K to $T^*$ is 
due to the slowing down of magnetic fluctuations often observed on cooling towards a magnetic transition. The drop of $a_0\eta$ 
is an indirect but perfectly clear indication of the onset of a spontaneous field at the muon site for a magnetic transition in a powder sample. 
Remarkably, there are anomalies neither in $a_0\eta(T)$ nor in $\lambda_Z(T)$ around $T_f$ = 16~K where the ZFC bulk susceptibility is maximum and shows a bifurcation with the FC susceptibility; see Fig.~\ref{muon_results}. Conversely there is no anomaly in the susceptibility at $T^*$. Furthermore, the temperature $T^*$ does not correspond to any of the broad maxima observed in $C_{M}/T$. For $T \ll T^*$, $a_{0}\eta$ is $\approx 45 \%$ of the value above the transition, relatively close to the value $1/3$ that would be expected for a powder sample with randomly-oriented local fields. 
As discussed previously, the early-time signal relaxes rapidly and is lost in the spectrometer dead time. (ii) The spin-lattice relaxation channel
is modeled by a stretched exponential function; the exponential relaxation expected in a homogeneous sample is only observed deep in the paramagnetic phase. This suggests that the compound is magnetically inhomogeneous: there is a distribution of spin-lattice relaxation rates \cite{Lindsey80,Berderan05,Johnston06,Yaouanc11}. (iii) Since an appreciable monotonic decay of the asymmetry is observed even at extremely low temperature, persistent spin dynamics are present. In fact $\lambda_Z(T)$ displays a plateau below $\approx$~2~K down to the lowest temperature (0.1~K)\@. This behavior seems to be a characteristics of frustrated magnetic compounds either with \cite{Yaouanc05a,Dalmas06,Chapuis09b} or without \cite{Keren00,Marcipar09} magnetic order. Interestingly, the presence of fluctuating iron moments was also deduced from the M\"ossbauer spectroscopy study of FeGa$_2$S$_4$ \cite{Myoung10}.

Quantitatively, we can 
fit the prediction of a model which treats nuclear/muon spin relaxation due to critical fluctuations of a 2D quantum frustrated Heisenberg antiferromagnet \cite{Chubukov94,Chubukov94a,Chubukov94b}, to $\lambda_Z(T)$ in the paramagnetic state.  
This model predicts for $T \ll T_0/2$
\begin{equation}
\lambda_Z \propto T^3 \exp(T_0/T),
\label{fit_extra}
\end{equation}
where $T_0$ is proportional to the spin-stiffness constant which is related to exchange.  Figure~\ref{muon_results} shows the best fit to the data. We find $T_0$ = 225\,(20) and 280\,(20)~K for the zero and 10~mT longitudinal field data, respectively. The exchange constant $J$ can be estimated from the relations $k_B T_0$ = $4\pi \rho_s$ \cite{Chubukov94a} and $\rho_s$ = $[1-0.399/(2S)]JS^2/\sqrt{3}$ for $S$ = 2 \cite{Chubukov94}: we find $J/k_B$ = 8.6\,(8) and 10.7\,(8)~K for the two values of $T_0$, respectively. They compare favorably with the value 13.3\,(7)~K derived from the Curie-Weiss temperature.

We have also recorded spectra after field cooling the sample in a $B_{\rm ext}$ = 10~mT longitudinal field; see Fig.~\ref{muon_spectra}. 
It appears that the application of the field strongly affects the spectral shape at short time. This can be interpreted as a signature of a change of the static or quasistatic field distribution at the muon site. 
At longer times the spectra reflect only the muon relaxation associated to the dynamics of the local field. Therefore, to gain an insight into these dynamics we fitted the experimental data after truncation of the early time part, using Eq.~(\ref{fit_1}). Cutoff times ranging between 100 and 300~ns were tested and the fitting parameters were found essentially independent of this cutoff. The results are shown in Fig.~\ref{muon_results}.
The field drastically suppresses $\lambda_Z$, and the temperature dependencies of the three parameters in the vicinity of $T^*$ 
are somewhat shifted to higher temperature than in zero field. 

The value of $B_{\rm ext}$ is much less than the width $\Delta B$ of the local field distribution to which the muons are submitted at low temperature. Therefore we would not expect the spectral shape at long time to be affected by $B_{\rm ext}$. This field effect is reminiscent of NiGa$_2$S$_4$ \cite{MacLaughlin08} where the internal field is much less distributed than in the current system, but still larger than 0.2~T at low temperature. 

\begin{figure}
\includegraphics[scale=0.75]{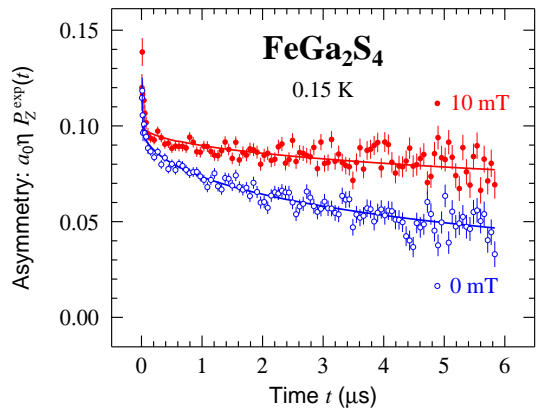}
\caption{(color online) Comparison of spectra measured at 0.15~K for FeGa$_2$S$_4$ after zero-field cooling. The first spectrum is measured in zero-field and the second after the subsequent application of a 10~mT longitudinal field.}
\label{muon_spectra_field}
\end{figure}

Additional measurements of the field response of FeGa$_2$S$_4$ have been performed at very low temperature. The 10~mT spectrum measured after a ZFC protocol and displayed in Fig.~\ref{muon_spectra_field}  again shows an important difference with the zero-field spectrum. Therefore the drastic effect of a modest field does not depend on the temperature of the field application and it is still present at 0.15~K.
This is in contrast to the field insensitivity of the specific heat for a field as large as 7~T \cite{Nakatsuji07b}.

The rapid increase of $\lambda_Z$ as the sample is cooled down in zero field towards $T^*$ is a key result of our study. In a conventional magnet it would indicate a magnetic phase transition, see for example Ref.~\onlinecite{Gubbens94a}. However, as no anomaly is observed in the magnetic specific heat and susceptibility at that temperature \cite{Nakatsuji07b,Nakatsuji10}, the detected anomaly in $\lambda_Z$ does not correspond to a conventional phase transition. Note that a rapid increase of $\lambda_Z$ as the sample is cooled down towards 9~K was also found for NiGa$_2$S$_4$ \cite{Yaouanc08,MacLaughlin08}.
The transition at $T^*$
is not spin-glass-like, since persistent spin dynamics are found even at 
temperatures as low as $0.1$~K\@.

As indicated earlier the NiGa$_2$S$_4$ transition at $T^*$ could be associated with a $Z_2$ topological transition. In the case of FeGa$_2$S$_4$ the transition which has been
revealed in the present study occurs at a temperature much higher than
those of the first broad peak in the magnetic specific heat 
($\approx$10~K) and of the bifurcation in the ZFC susceptibility ($\approx$16~K).
Thus it seems unlikely that the transitions in FeGa$_2$S$_4$ and NiGa$_2$S$_4$ are of the same nature.
This conclusion is further supported by the fact that hysteresis effects
for the latter compound 
are much smaller and occur at $T^*$.  

In conclusion, our $\mu$SR experiments on FeGa$_2$S$_4$ 
confirm the presence of the magnetic transition at $T^* \simeq 30$~K\@ in zero 
field which was previously observed by $^{57}$Fe M\"ossbauer spectroscopy.
With the data available we are unable to estimate the
magnitude of the magnetic moment involved. Because of the 
lack of anomalies at $T^*$ in the specific heat and in the susceptibility and because of a strong field effect, this transition is exotic.
In addition, as usual for geometrically frustrated magnetic compounds and in particular for NiGa$_2$S$_4$, persistent spin dynamics are observed at low temperature. 

This work was supported in part by the European Science Foundation through the Highly Frustrated Magnetism program, by the U.S. National Science Foundation, Grant No.~0801407, and by the Japan MEXT, Grants-in-Aid Nos.~17071003 and 19052003. The $\mu$SR experiments were performed at the Swiss Muon Source, Paul Scherrer Institute, Villigen, Switzerland.

\bibliography{reference.bib}

\end{document}